\documentclass[sort&compress,twocolumn,5p]{elsarticle}
\pdfoutput=1

\usepackage{amsmath}

\usepackage[breaklinks,
            pdftitle={Computational Aspects of Speed-Dependent Voigt and Rautian Profiles},
            pdfsubject={Approximations and expansions, Computational techniques},
            pdfkeywords={Complex error function, Voigt function; Speed-dependent Voigt function; Speed-dependent Rautian function; Rational approximations},
            pdfauthor={F. Schreier and P. Hochstaffl}]{hyperref}

\newcommand{\qeq}[1]  {Eq.~(\ref{#1})}
\newcommand{\qufig}[1] {Fig.~\ref{#1}}
\newcommand{\qutab}[1] {Tab.~\ref{#1}}
\newcommand{\D}{\mathrm{d}}
\newcommand{\I}{\mathrm{i}}
\newcommand{\erfc} {\mathop{\rm erfc}}
\newcommand{\sign} {\mathop{\rm sign}}
\newcommand{\chem}[1] {{\ensuremath{\mathrm{#1}}}}
\renewcommand{\Re} {\mathop{\rm Re}}

\batchmode

\journal{JQSRT, received 27 August 2020, accepted 10 October 2020; doi: 10.1016/j.jqsrt.2020.107385}

\begin{document}

\begin{frontmatter}

\title{Computational Aspects of Speed-Dependent Voigt and Rautian Profiles}

\author{Franz Schreier\corref{fs}}
\ead{franz.schreier@dlr.de}
\cortext[fs]{Corresponding author} 
\author{Philipp Hochstaffl\corref{ph}}
\address{DLR --- German Aerospace Center, 
         Remote Sensing Technology Institute, \\
         82234 Oberpfaffenhofen, Germany}

\begin{abstract}
For accurate line-by-line modeling of molecular cross sections several physical processes ``beyond Voigt'' have to be considered.
For the speed-dependent Voigt and Rautian profiles (SDV, SDR) and the Hartmann-Tran profile the difference $w(\I z_-)-w(\I z_+)$ of two complex error functions (essentially Voigt functions) has to be evaluated where the function arguments $z_\pm$ are given by the sum and difference of two square roots.
These two terms describing $z_\pm$ can be huge and the default implementation of the difference can lead to large cancellation errors.
First we demonstrate that these problems can be avoided by a simple reformulation of $z_-$.
Furthermore we show that a single rational approximation of the complex error function valid in the whole complex plane (e.g.\ by Huml\'\i\v{c}ek, 1979 or Weideman, 1994) enables computation of the SDV and SDR with four significant digits or better.
Our benchmarks indicate that the SDV and SDR functions are about a factor 2.2 slower compared to the Voigt function,
but for evaluation of molecular cross sections this time lag does not significantly prolong the overall program execution because speed-dependent parameters are available only for a fraction of strong lines.
\end{abstract}



\end{frontmatter}

\renewcommand{\thefootnote}{\fnsymbol{footnote}}
\footnotetext{This manuscript version is made available under the CC-BY-NC-ND 4.0 \url{http://creativecommons.org/licences/by-nc-nd/4.0/}}
\renewcommand{\thefootnote}{\arabic{footnote}}

\section{Introduction}
\label{sec:intro}

The Voigt profile \citep{Armstrong67} accounting for collision (pressure) and Doppler (thermal) broadening has been the standard for high-resolution line-by-line modeling of infrared (IR) and microwave
molecular absorption.
Inadequacies of this profile have been observed since decades in molecular laboratory spectroscopy \citep[e.g.][]{Dicke53},
and more recently discrepancies between theory and measurements have also become significant in atmospheric spectroscopy.

Collision induced changes of molecular velocity reduce the Doppler broadening, and several profiles have been suggested to describe this collisional (or Dicke) narrowing, e.g.\ the Rautian
profile \citep{Varghese84}.
Moreover, the speed-dependence of the relaxation rates modifies the Lorentz line shape modeling the collision broadening \citep{Hartmann08,Tennyson14} and can be described by the Speed-Dependent
Voigt (SDV) profile.
Assuming that these processes are independent leads to the Speed-Dependent Rautian (SDR) profile.
The partial correlation of these effects is considered by the partially Correlated quadratic-Speed-Dependent Hard-Collision profile (pCqSDHCP),
originally developed by \citet*{Tran13} (therefore called Hartmann-Tran or HT profile for short) and recommended for high resolution spectroscopy in a recent IUPAC Technical Report \citep{Tennyson14}.

Using these more sophisticated profiles has been shown to improve the analysis of ground-based Fourier transform IR spectra and space-borne limb occultation measurements of the Atmospheric
Chemistry Experiment --- Fourier Transform Spectrometer (ACE-FTS) \citep[e.g.][]{Schneider09b,Schneider11,Boone07}.
The stringent accuracy and precision requirements of current space-borne carbon dioxide and methane observations (e.g.\ the Greenhouse Gases Observing Satellite \citep[GOSAT,][]{Kuze09},
the Orbiting Carbon Observatory \citep[OCO-2,][]{Crisp04} and the Sentinel-5p TROPOspheric Monitoring Instrument \citep[TROPOMI,][]{Veefkind12}) have also highlighted the need for improved
molecular spectroscopy in the short-wave IR \citep[e.g.][]{Nikitin10,Nikitin15,Oyafuso17,Galli14,ChecaGarcia15,Hochstaffl20s,Hochstaffl20t}.
Improved Atmospheric Spectroscopy databases (IAS) have been the objective of a study funded by ESA in the context of the Scientific Exploitation of Operational Missions (SEOM) initiative:
SEOM-IAS provides data for \chem{CO}, \chem{CH_4} and \chem{H_2O} in the TROPOMI $2.3\rm\,\mu m$ region \citep{Birk17}.
Moreover, the HITRAN database has started to include ``beyond-Voigt'' parameters for selected molecules \citep{Gordon17etal}.

Computational aspects of the SDV profile have been discussed in a previous paper \citep{Schreier17} (henceforth ``CASDV'').
Here we extend this discussion to the SDR profile.
After a brief survey of available line data we recall some definitions and relations in the following section.
We continue with a discussion of some subtleties of the SDV function and then present comparisons of SDR evaluations against a reference code in section \ref{sec:results}.
Moreover, we provide an extensive assessment of the computational efficiency of the SDV and SDR (floating-point operation counts and time benchmarks).
A summary is given in the final section \ref{sec:conclusions}.
For simplicity, pressure-induced line shift, self-broadening, and line mixing will be ignored.
Python modules of the SDV and SDR functions along with rational approximations for the complex error function are provided in supplemental files.


\section{Line data and models}
\label{sec:theory}

\subsection{Line data}
\label{sec:data}
For an assessment of the performance of various line model algorithms and implementations it is useful to recap the range of the function arguments to be expected.
In \citet{Schreier11v, Schreier18h} we concluded that the ratio of the Lorentz and Gauss width (the $y$ parameter, see next subsection) can be as small as $10^{-8}$ for atmospheric IR spectroscopy.
Even smaller values are possible for large wavenumbers (visible or UV), light molecules (e.g.\ \chem{H_2}) and warm to hot atmospheres.

\begin{figure*}
 \centering\includegraphics[width=\textwidth]{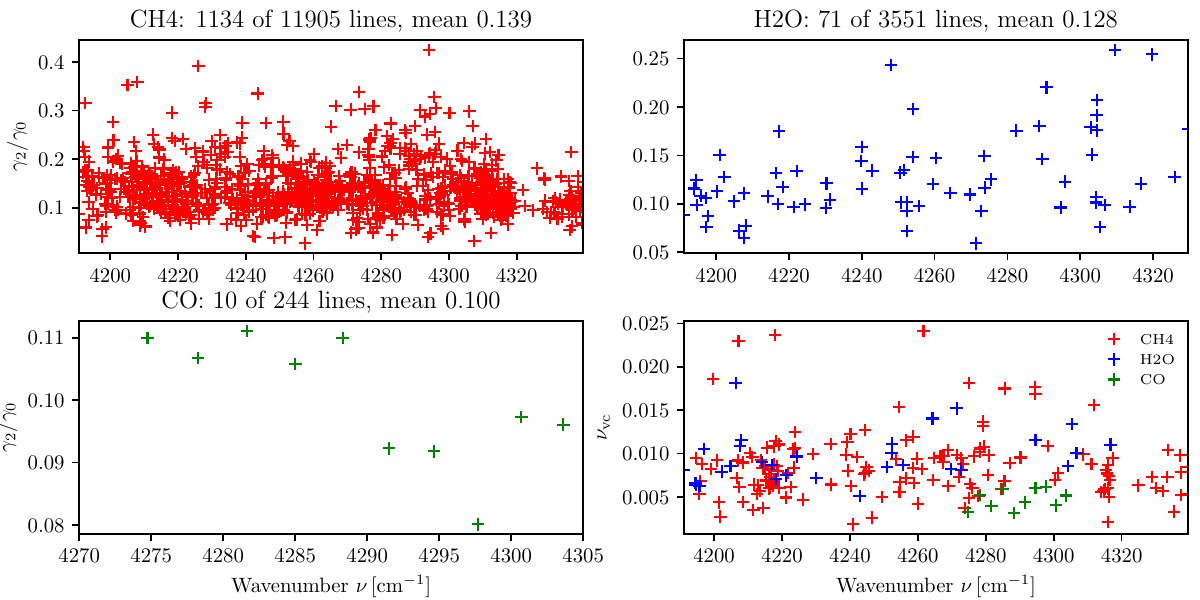}
 \caption{Survey of SEOM-IAS line data.
          Top and bottom left: ratio of the collision broadening parameters, $\gamma_2/\gamma_0$
          (the title indicates the number of lines with nonzero $\gamma_2$ and the mean value of the ratio).
          Bottom right: Dicke narrowing parameter $\nu_\text{vc}$.}
 \label{fig:seom} 
\end{figure*}

Considering the new parameters required by the speed-dependent profiles, a survey of the SEOM-IAS data indicates that the ratio of the collision broadening parameters has a mean value of
$\gamma_2/\gamma_0 \approx 0.1$, see \qufig{fig:seom}.
Furthermore, the bottom-right plot of \qufig{fig:seom} shows that the frequency of velocity-changing collisions (Dicke narrowing) is about one tenth of the air-broadening parameter,
i.e.\ $\nu_\text{vc} \approx 0.01\rm\,cm^{-1} \approx 0.1 \gamma_0$.
Note that in SEOM-IAS the partial correlation between speed-dependence and velocity changes is not considered, i.e.\ the HT parameter $\eta$ is assumed to be zero and the SDR is sufficient for this database.
Furthermore, the line shift's speed dependence parameter $\delta_2$ is zero for all carbon monoxide and water lines, whereas 12 of the 11905 methane lines have nonzero values with $-0.006545 \le \delta_2 \le 0.00316 \rm\,cm^{-1}$.

\begin{figure*}
 \centering{\includegraphics[width=\textwidth]{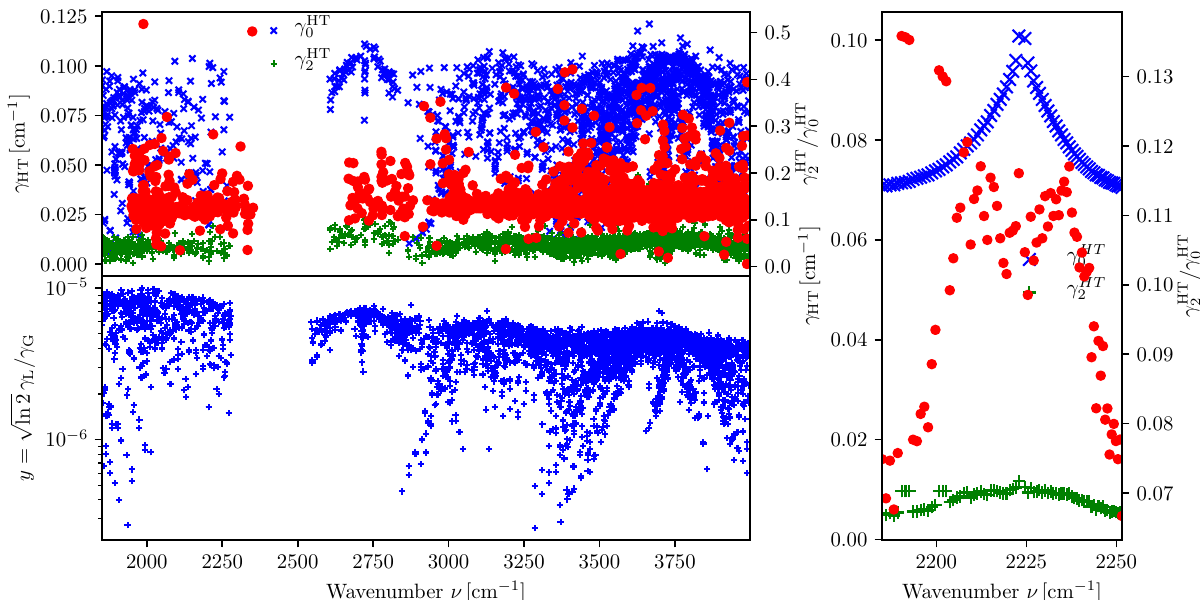}}
 \caption{HITRAN \chem{H_2O} (left) and \chem{N_2O} (right) lines with HT widths defined:
          Top left: collision broadening parameters (x and + markers, left axis) and their ratio $\gamma_2/\gamma_0$ (red filled circles, right axis).
          Bottom left: the Voigt parameter $y$ for ToA @ 120\,km.
          }
 \label{fig:hitran}
\end{figure*}

In the latest version of the HITRAN database, ``beyond-Voigt'' parameters are available for seven (of 49) molecules according to Table 3 of \citet{Gordon17etal}.
However, these are not necessarily HT or SDV parameters and might describe, for example, line mixing.
For \chem{H_2O} there are 2016 (of about 52\,000) lines with nonzero $\gamma_2$ in the spectral range $1852 \mbox{\,--\,} 3995 \rm\,cm^{-1}$ and a minimum line strength $S = 6.9 \cdot 10^{-25} \rm\, cm^{-1} / (molec.cm^{-2})$, see \qufig{fig:hitran} (left) and \citet{Loos17a,Loos17p};
moreover the Dicke narrowing parameter is nonzero for 237 transitions with a mean of $0.0157\rm\,cm^{-1}$.
For \chem{N_2O} 80 lines around $2200\rm\, cm^{-1}$ with $S \ge 10^{-19} \rm\, cm^{-1} / (molec.cm^{-2})$ are listed (\qufig{fig:hitran}).
For both molecules the speed dependence of the line width is about one tenth of the speed-averaged width, and the correlation parameter $\eta$ is zero for all transitions.
(For \chem{H_2} see also \citep{Wcislo16}.)

Similar to HITRAN and GEISA, the SEOM-IAS line parameters are given for a reference pressure $p_\text{ref}=1\rm\,atm$ and temperature $T_\text{ref}=296\rm\,K$.
Both broadening parameters are assumed to have the same pressure and temperature dependence, i.e.\ $\gamma_{0,2} \sim p T^{-n}$ with the temperature exponent $n$ ($1/2$ according to classical theory), whereas the narrowing parameter is inversely proportional to temperature, $\nu_\text{vc} \sim p/T$
\citep[see also][]{Gamache20,Stolarczyk20}.
In Earth's (or a planetary) atmosphere these three parameters are varying over orders of magnitude, but the ratios $\gamma_2/\gamma_0$ and $\nu_\text{vc} / \gamma_0$ are approximately constant over pressure (or altitude).

\subsection{Line models}
\label{sec:models}

Following \citet{Varghese84} we distinguish between the Voigt profile $g$ and its generalizations 
and the Voigt function $K$ and its generalizations 
\begin{equation} \label{vgtFct} 
 g(\nu;\hat\nu,\gamma_\text{L},\gamma_\text{G}, \dots) ~=~ {\sqrt{\ln 2 / \pi} \over \gamma_\text{G}} \, K(x,y, \dots) ~.
\end{equation}
The profiles $g$ are defined in terms of the physical variables wavenumber $\nu$, center position $\hat\nu$, and broadening parameters, all in units of reciprocal wavenumbers $[\rm cm^{-1}]$
(in the following all $\gamma$ as well as $\nu_\text{vc}$ are considered for the actual pressure and temperature).
Division by the Gaussian (Doppler) half width allows to introduce dimensionless variables that reduces the number of arguments by one and is more convenient from a mathematical point:
\begin{subequations}
 \begin{alignat}{2}
 \label{voigtXY} x &\equiv \sqrt{\ln 2} (\nu-\hat\nu)/\gamma_\text{G}  & \qquad  y &\equiv \sqrt{\ln 2}\gamma_\text{L}/\gamma_\text{G} \\
 \label{sdVoigt} q &\equiv \sqrt{\ln 2}\gamma_\text{2}/\gamma_\text{G} & \qquad  \zeta &\equiv \sqrt{\ln 2}\nu_\text{vc}  /\gamma_\text{G} ~.
 \end{alignat}
\end{subequations}

From a computational point, these profiles and corresponding functions can be calculated readily from the complex error function \citep{AbSt64,NIST-handbook,DLMF}
\begin{align} \label{cerf}
 w(z)   ~&=~ {\I \over \pi} ~ \int_{-\infty}^\infty {\mathrm{e}^{-t^2} \over z-t} \, \D t \\
         &=~ K(x,y) + \I L(x,y)   \qquad\text{with $z=x+\I y$.}
\end{align}
The real part is the ``standard'' Voigt function, essentially the convolution of a Lorentzian and Gaussian function, 
\begin{equation}\label{vgtFct}
 K(x,y) ~=~ {y \over \pi} ~ \int_{-\infty}^\infty {\mathrm{e}^{-t^2} \over (x-t)^2 + y^2} \, \D t ~,
\end{equation}
where $x$ and $y$ are the standard Voigt function parameters \eqref{voigtXY} characterizing the distance to the line center at wavenumber $\hat\nu$ and the ratio of the Lorentzian and Gauss widths.

The SDR function is essentially the quotient of the difference of two complex error functions
\begin{equation}
\label{sdrFunction}
 K_\text{sdr}(x,y,q,\zeta) ~=~  \Re\left( {w(\I z_-) - w(\I z_+) \over 1 - \sqrt\pi \zeta \Bigl(w(\I z_-) - w(\I z_+)\Bigr)} \right)
\end{equation}
where the extra parameters $q$ and $\zeta$ are the broadening's quadratic speed dependence $\gamma_2$ and the frequency of velocity-changing collisions $\nu_\text{vc}$ normalized by the Gaussian widths, respectively, \qeq{sdVoigt}.
The arguments of the complex error function are defined by
\footnote{The upper case $X$ and $Y$ correspond to the HT notation \citep{Tennyson14,Tran13} with 
$X \equiv \bigl[(\I(\hat\nu-\nu) + \gamma_\text{L}+\nu_\text{vc} \bigr] / \gamma_2 - {3 \over 2}$ 
and $Y \equiv \left({\gamma_\text{G} \big/ 2 \sqrt{\ln 2} \gamma_2}\right)^2$.
Note that $Y$ is real, a non-vanishing line shift speed-dependence parameter would lead to a complex $Y$.
\citet{Boone11} define two variables $\alpha$ and $\beta$ corresponding to the real and imaginary part of $X$ (with $\nu_\text{vc}=0$), a third variable $\delta$ is identical to $Y$.
The function $\Re(A)$ defined in Eq.\ (7) of \citet{Tennyson14} is identical to the ``SDV function'' $K_\text{sdv}$ except for a factor $\sqrt{\pi \ln 2} / \gamma_\text{G}$.
Note that in footnote 1 of \citep{Schreier17} we have incorrectly stated that an Im operator is missing in Eq.\ (11) of \citet{Tran13}.}
\begin{align}
\label{sdrZ}    z_\pm ~&=~ \sqrt{{y + \zeta - \I x \over q} - {3 \over 2} + {1 \over 4q^2}} ~\pm~ {1 \over 2q} \\
\label{sdrZtnh}        &=~ \sqrt{X+Y} ~\pm~ \sqrt{Y} ~.
\end{align}
For vanishing $\nu_\text{vc} \propto \zeta = 0$ the SDR function reduces to the SDV function
\begin{equation} \label{sdvFunction}
K_\text{sdv}(x,y,q) ~=~  \Re\bigl( w(\I z_-) - w(\I z_+) \bigr),
\end{equation}
whereas for zero speed dependence $\gamma_2 \propto q = 0$ the Rautian function is obtained,
\begin{equation} \label{rautian}
K_\text{r}(x,y,\zeta) ~=~   \Re\left( {w(z) \over 1 - \sqrt\pi \zeta w(z)} \right) ~.
\end{equation}

A vast number of algorithms have been developed for the complex error function exploiting a wide variety of numerical techniques.
In recent papers we have discussed rational approximations by \citet{Humlicek79,Humlicek82} and \citet{Weideman94} and recommended combinations of the asymptotic approximation given in the \citet{Humlicek82} \texttt{w4} code (the quotient of a first and second degree polynomial) with the more accurate approximations of \citep{Humlicek79} or \citep{Weideman94} for small arguments \citep{Schreier11v,Schreier18h}.


\section{Results}
\label{sec:results}

\subsection{Computation of the complex square roots}
\label{sec:arguments}

The arguments $\I z_\pm$, \qeq{sdrZ}, of the complex error functions for \eqref{sdrFunction} and \eqref{sdvFunction} require the computation of a complex square root.
In essence\footnote{This is essentially the square root used by \citet{Boone11} without the $\delta$ term.},
\begin{subequations} \label{zpm}
\begin{align}
 \label{zPMtnh}
 r     ~&=~ \sqrt{\alpha +  \I \beta} \\
 \label{zPMbwb}
       ~&=~ {1 \over \sqrt 2} \sqrt{\sqrt{\alpha^2 + \beta^2} +\alpha} ~+~ \sign(\beta) {\I \over \sqrt 2} \sqrt{\sqrt{\alpha^2 + \beta^2} -\alpha}  \\
 \label{zPMnew}
       ~&=~ {1 \over \sqrt 2} \sqrt{\sqrt{\alpha^2 + \beta^2} +\alpha} ~+~ {\I \beta / \sqrt 2 \over \sqrt{\sqrt{\alpha^2 + \beta^2} +\alpha }}
\end{align}
\end{subequations}
The three variants of \eqref{zpm} are all equivalent mathematically, but might be different numerically.

\citet*{Tran13} (henceforth ``TNH'') evaluate the square root of the complex argument using Fortran's intrinsic function \texttt{CDSQRT}, essentially the right hand side of  \qeq{zPMtnh},
whereas the \citet*{Boone11} (henceforth ``BWB'') Fortran implementation is based on \qeq{zPMbwb}.
In CASDV we have indicated that the alternative form \eqref{zPMnew} might be advantageous computationally because possible inaccuracies of the difference evaluation can be avoided.

With respect to efficiency, it requires only two instead of three square root evaluations (but an extra division) and there is no need to evaluate the sign function.
Note that the Voigt and generalized Voigt functions are symmetric w.r.t.\ $x \propto \beta \equiv \mathrm{Im}(X)$, i.e.\ they depend on $|\beta|$ only and the \texttt{sign} function can be avoided.

\begin{figure*}[h]
 \centering\includegraphics[width=\textwidth]{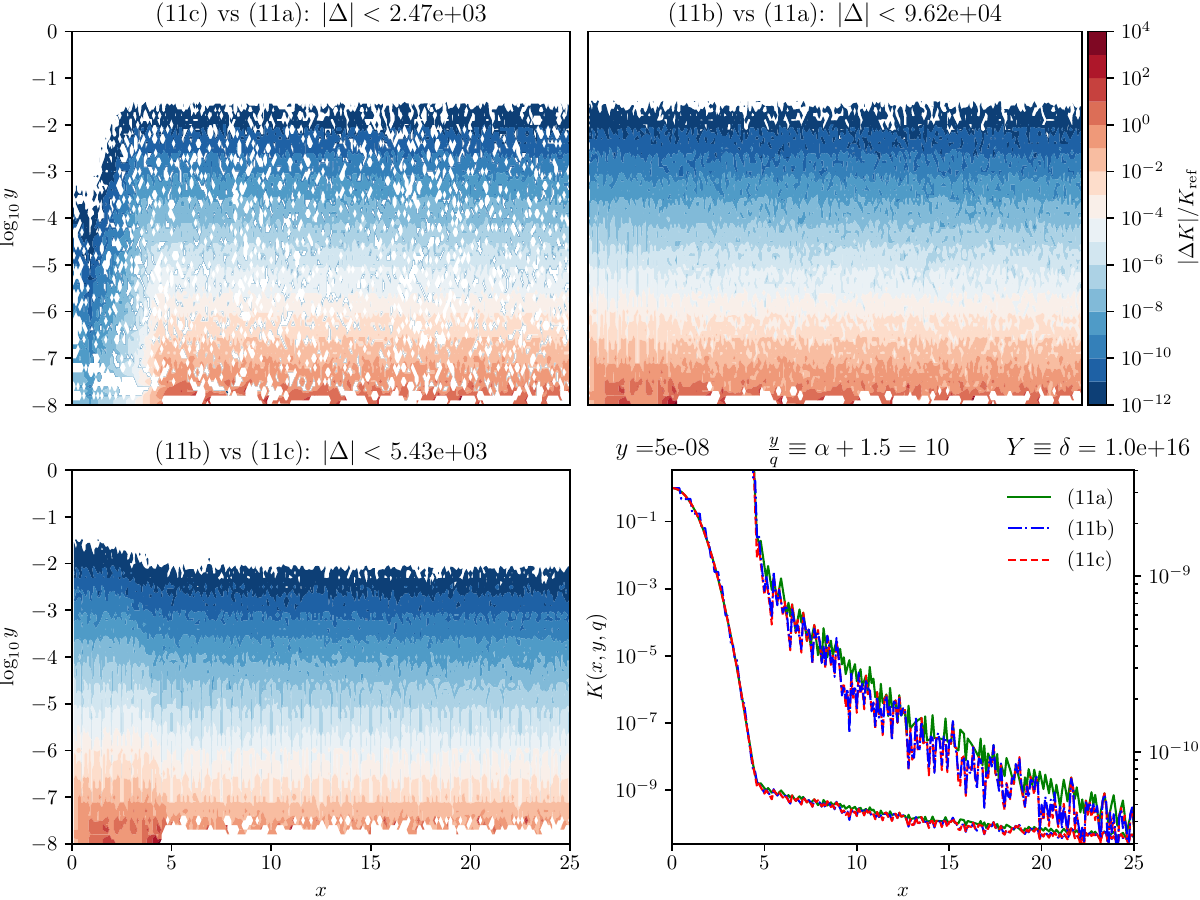}
 \caption{Contour plot of relative differences of the SDV function for $\gamma_\text{L}/\gamma_2=10$: mutual differences of the three versions shown in \qeq{zpm}.
          Bottom right: the SDV functions for $y=5\cdot 10^{-8}$; the curves are plotted twice with the entire $K$ range (left axis) and with the $K$ range limited to small values (right axis).}
 \label{fig:sdv_z} 
\end{figure*}

The contour plots of \qufig{fig:sdv_z} demonstrate that the three forms are equivalent except for very small $y$, i.e.\ the relative differences $|\Delta K| / K$ are small.
However, discrepancies can be observed for $y<10^{-6}$.
Inspection of the SDV for $y=5\cdot 10^{-8}$ (\qufig{fig:sdv_z} bottom right) indicates that \eqref{zPMbwb} produces discrete steps near the line center (small $x$), where the other two forms are indistinguishable.
In the wings (larger $x$) the three forms show a zigzag behaviour; \eqref{zPMbwb} and \eqref{zPMnew} are quite similar and somewhat smaller than \eqref{zPMtnh} that oscillates around the Voigt function.
Fortunately the function values here are many orders of magnitude smaller compared to the center value and the apparent errors are likely to have no impact on molecular cross sections resulting from the superposition of numerous lines.
Note that TNH carefully discuss limiting asymptotic cases and have implemented dedicated algorithms for small and large $|X|/Y$ in the supplemental Fortran source code.

\subsection{Problems with tiny values of \texorpdfstring{$y$ and $q$}{y and q}.}
\label{sec:yTiny}
Although the direct evaluation of the complex square \eqref{zPMtnh} appears to be better, the zigzag indicates further problems.
A closer analysis indicates that these problems are related to the evaluation of $z_-$, i.e.\ the difference of the complex square root $\sqrt{X+Y}$ and $\sqrt{Y}$.
For tiny $y$ (and $q<y$) the real valued $Y=1/4q^2$ becomes extremely big, much bigger than the complex valued $X$, and the standard floating point precision with about 16 significant digits fails to evaluate this difference.
To test this hypothesis, the SDV has been implemented using the Python library \texttt{mpmath} \citep{mpmath} allowing real and complex floating-point arithmetic with arbitrary precision.
With 32 significant figures (roughly quadruple precision) the square roots, their difference and the complex error function (computed by $w(z) = \exp(-z^2)\erfc(-\I z)$) can be evaluated reliably even for these extreme cases.

\begin{figure*}[h]
 \centering\includegraphics[width=\textwidth]{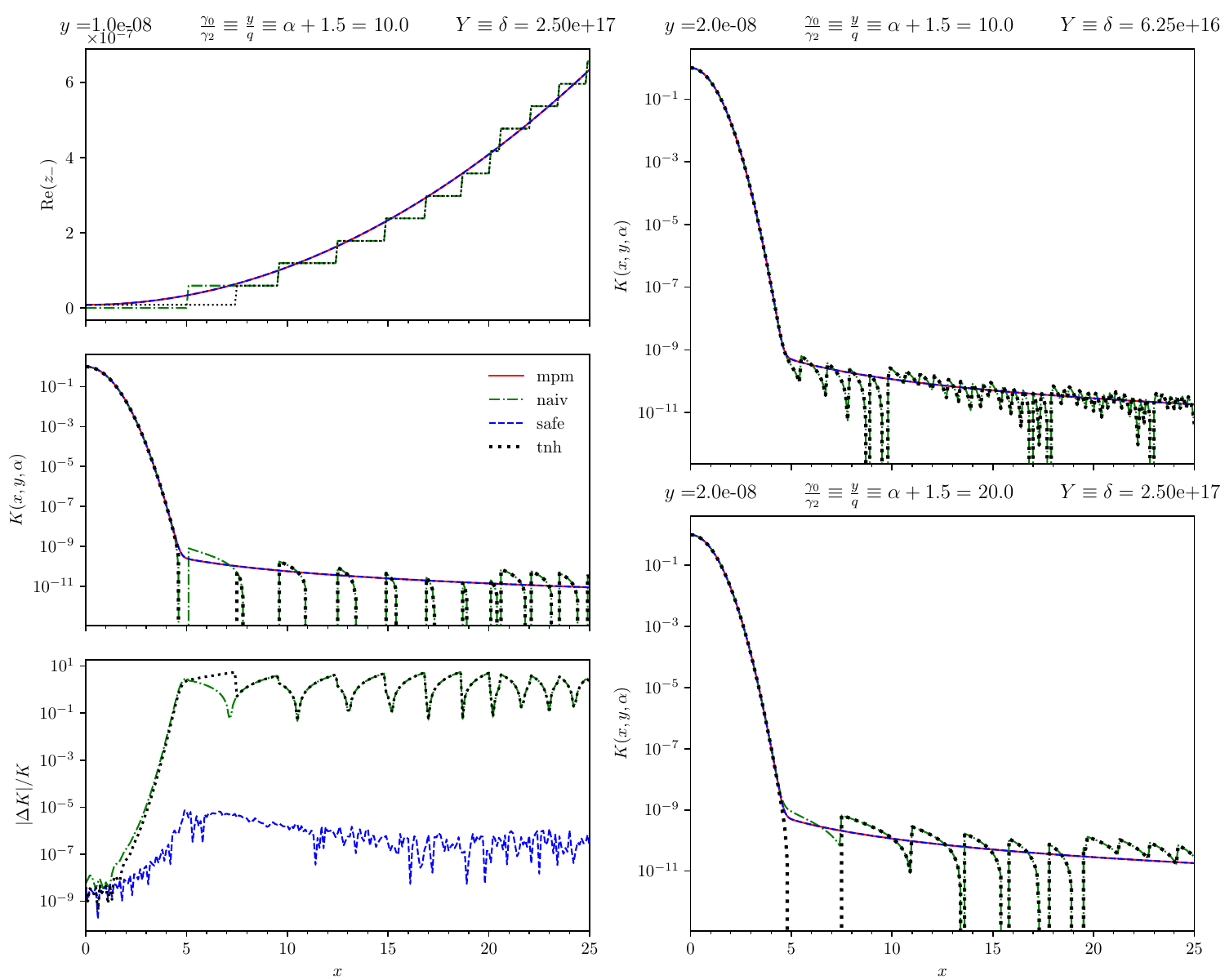}
 \caption{The SDV function evaluated for tiny values of $y$.
          Top left: the real part of the complex error function argument $z_-$ for $y=1\cdot 10^{-8}$ and $q=1\cdot 10^{-9}$.
          Center left: the corresponding SDV function values.
          Bottom left: the relative error w.r.t.\ the ``mpm'' quadruple precision evaluation.
          Right: the SDV functions for $y=2\cdot 10^{-8}$ and two values of $q$.}
 \label{fig:sdv_yTiny} 
\end{figure*}

\qufig{fig:sdv_yTiny} (left) shows results for $y=10^{-8}$ and $q = 0.1 y$ (some specific numbers for $x=10$ and $x=12$ are listed in the appendix):
Whereas $\alpha={y \over q} - {3 \over 2}$ is a small number, $Y \equiv \delta = {1 / 4 q^2}$ is huge, and with the default double precision accuracy of Python (or C or Fortran) the real part of the first square root in \eqref{zpm} is essentially independent of $\alpha$.
For the $x$ values considered here ($0 \le x \le 25$) the imaginary part of $X+Y$, i.e.\ ${x/q}$ is a moderately large number.
However, the real part of the complex square root $\sqrt{X+Y}$ is identical in double precision for $x=10$ and $x=12$ (but clearly distinct with enhanced precision).
Moreover, because $|X| \ll Y$, the difference of two huge, approximately similar numbers cannot be evaluated correctly,
resulting in a stepwise behaviour of $\mathrm{Re}\left(\sqrt{X+Y}-\sqrt{Y}\right)$ (\qufig{fig:sdv_yTiny} top-left).
As a consequence, the evaluation of $w(\I z_-)$ is unreliable, too;
in particular, the difference of the complex error functions $w(\I z_-) - w(\I z_+)$ can become negative, indicated by the missing values in \qufig{fig:sdv_yTiny} (center-left).

Similar to the reformulation discussed above (Eqs.\ \eqref{zPMbwb} and \eqref{zPMnew}) the difference of the two roots \eqref{sdrZtnh} can also be computed in a reliable way to avoid the subtraction of two similar numbers \citep{Panchekha15}\footnote{See also \url{https://herbie.uwplse.org/}},
\begin{equation} \label{zMinus}
 z_- ~\equiv~ \sqrt{X+Y} - \sqrt{Y} ~=~ {X \over \sqrt{X+Y} + \sqrt{Y}} ~=~ {X \over z_+}~.
\end{equation}
\qufig{fig:sdv_yTiny} shows that this safe variant avoids the steps in $\Re(z_-)$ and gives SDV values close to the quadruple precision values;
whereas the relative error of the original version is in the percentage range, the cancellation-safe implementation has a maximum relative error of about $10^{-5}$ (bottom-left).
The superiority of this approach is also confirmed by the two examples shown on the right of \qufig{fig:sdv_yTiny}, where the original ``naive'' computation delivers oscillations with frequent negative SDV values.

\subsection{The SDV function again}
\label{sec:sdv}

Both TNH and BWB have emphasized that the computation of the SDV and HT profiles needs to be done carefully when different complex error function approximations are used for the two arguments $\I z_\pm$.
In particular, the TNH code implements the \citet{Humlicek79} \texttt{cpf12} subroutine (with about $10^{-6}$ maximum relative error) that utilizes two different approximations for large and small arguments (in fact augmented with a 15-term asymptotic expansion of $w(z)$ for $|z|>8$).
When the two arguments are close to each other, TNH ensure that they are evaluated with the same method.
BWB exploit the \citet{Humlicek82} \texttt{w4} code ($10^{-4}$ relative accuracy) with four different approximations:
their code uses the higher degree rational approximation for both arguments.
However, as discussed in CASDV, this approach is dangerous because the higher degree approximation is not necessarily valid in the region of large function arguments.

Obviously, these problems could be avoided if a complex error function algorithm is used that is valid for the entire $x,y$ range.
\citet{Weideman94} has developed rational approximations that can be used in the entire complex plane. 
As demonstrated in \citep{Schreier11v}, the $n=24$ approximation has a maximum relative error less than $10^{-4}$ for $y>10^{-4}$ only;
with $n=32$ terms the code is better than $9 \cdot 10^{-5}$ for all $0\le x \le 20$ and $10^{-6} \le y \le 10^2$ for both the real and imaginary part of $w$. 
The region I approximation of the \citet{Humlicek79} \texttt{cpf12} code cannot be used for $y < 10^{-2}$, but a generalization of this 12-term rational approximation applicable to all $(x,y)$ considerably improves
the performance \citep{Schreier18h}: the 16-term approximation (\texttt{cpf16}) achieves an accuracy of about $10^{-4}$ for $y>10^{-8}$,
and with 20 terms (\texttt{cpf20}) the error can be further reduced by more than a factor ten.

\begin{figure*}[t]
 \centering\includegraphics[width=\textwidth]{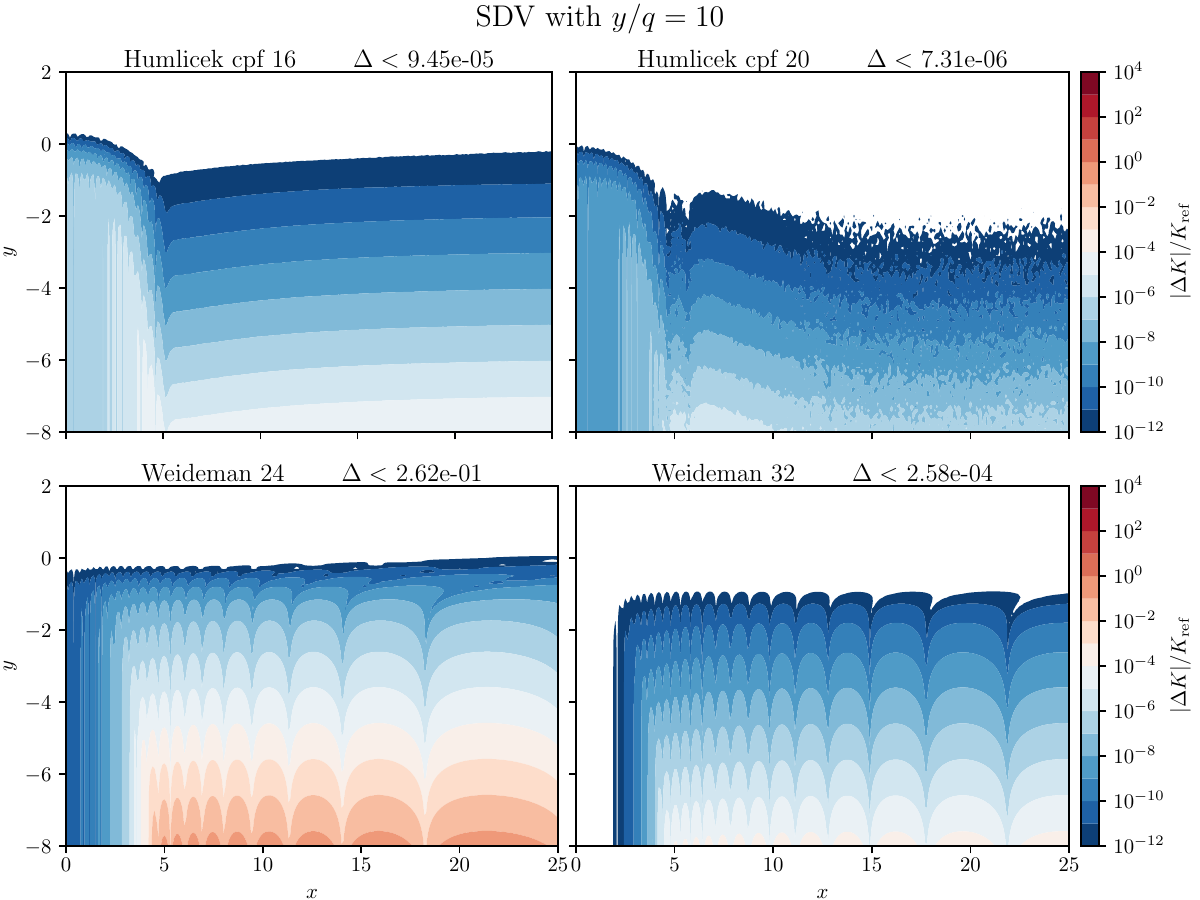}
 \caption{Contour plot of relative differences of the SDV function computed with the \protect\citeauthor{Humlicek79} (top) and \protect\citeauthor{Weideman94} complex error function codes for $\gamma_\text{L}/\gamma_2=10$.
          The number in the title indicates the maximum relative error.}
 \label{fig:cerf4sdv} 
\end{figure*}

\qufig{fig:cerf4sdv} depicts the relative errors of the SDV computed with these ``global'' complex error function algorithms (the 16- and 20-term \citet{Humlicek79} and the 24- and 32-term \citet{Weideman94} approximations) for $\gamma_\text{L}/\gamma_2 = y/q = 10$.
As in our previous works we use the \texttt{wofz} complex error function provided by SciPy (module \texttt{scipy.special}) with a stated accuracy of 13 significant digits as a reference code \citep[see also][]{faddeeva,libcerf}.
All implementations work reasonably good for $y>10^{-4}$, but for smaller $y$ problems become apparent especially for the Weideman $n=24$ approximation.
The Weideman $n=32$ and the \citeauthor{Humlicek79} 16-term approximation can be used safely for $y>10^{-6}$,
but only \texttt{cpf20} is able to evaluate the SDV with errors less than $5 \cdot 10^{-5}$ for $0\le x \le 25$ and $10^{-8} \le y \le 10^2$.
Note that the error patterns for the SDV deviations are similar to the pattern of the complex error function deviations
(cf.\ Fig.\ 7 in \citep{Schreier11v} and Fig.\ 2 and 4 in \citep{Schreier18h}).

\subsection{Beyond-Voigt vs.\ Voigt}
\label{sec:sdv-sdr}

\begin{figure*}[t]
 \centering\includegraphics[width=\textwidth]{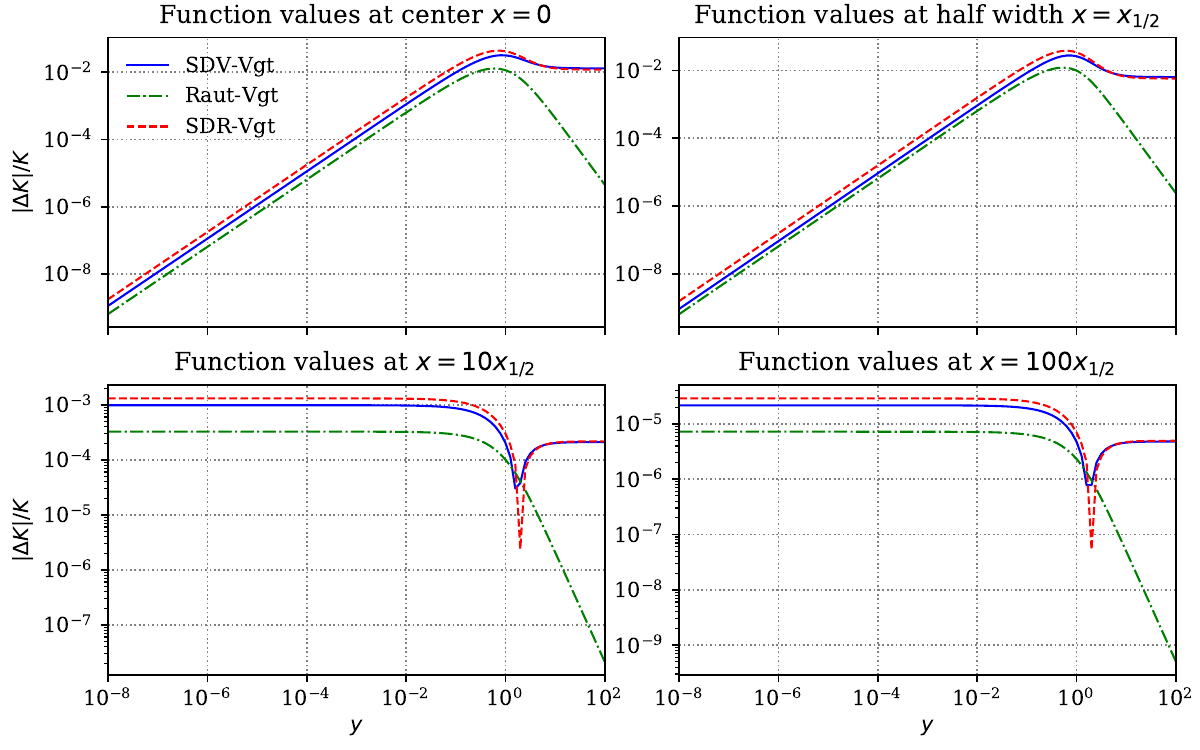}
 \caption{Relative differences of the SDR, SDV, and Rautian relative to the Voigt function as a function of $y$ with $y/q = y/r = 0.1$.}
 \label{fig:vgt-sdr} 
\end{figure*}

Before analyzing the performance of various complex error function algorithms for the evaluation of the SDR it is instructive to see the relative change of the function due to the inclusion
of velocity-changing collisions.
Figure \ref{fig:vgt-sdr} shows that the relative change $|K_\text{sdr}-K_\text{vgt}| / K_\text{vgt}$ can be as large as some percent.
Differences between SDR and Voigt up to about 4\% can be observed in the line center region for $y \approx 1$, whereas for large $x \gg x_{1/2}$ the relative difference is less than $10^{-4}$, i.e.\ the advanced profiles essentially resemble the Voigt profile asymptotically.
Here the Voigt function half width is given by the \citet{Whiting68} approximation $x_{1/2} = \tfrac{1}{2} \left(y+\sqrt{y^2 + 4 \ln 2} \right)$.


\subsection{The SDR function}
\label{sec:sdr}

\begin{figure*}[t]
 \centering\includegraphics[width=\textwidth]{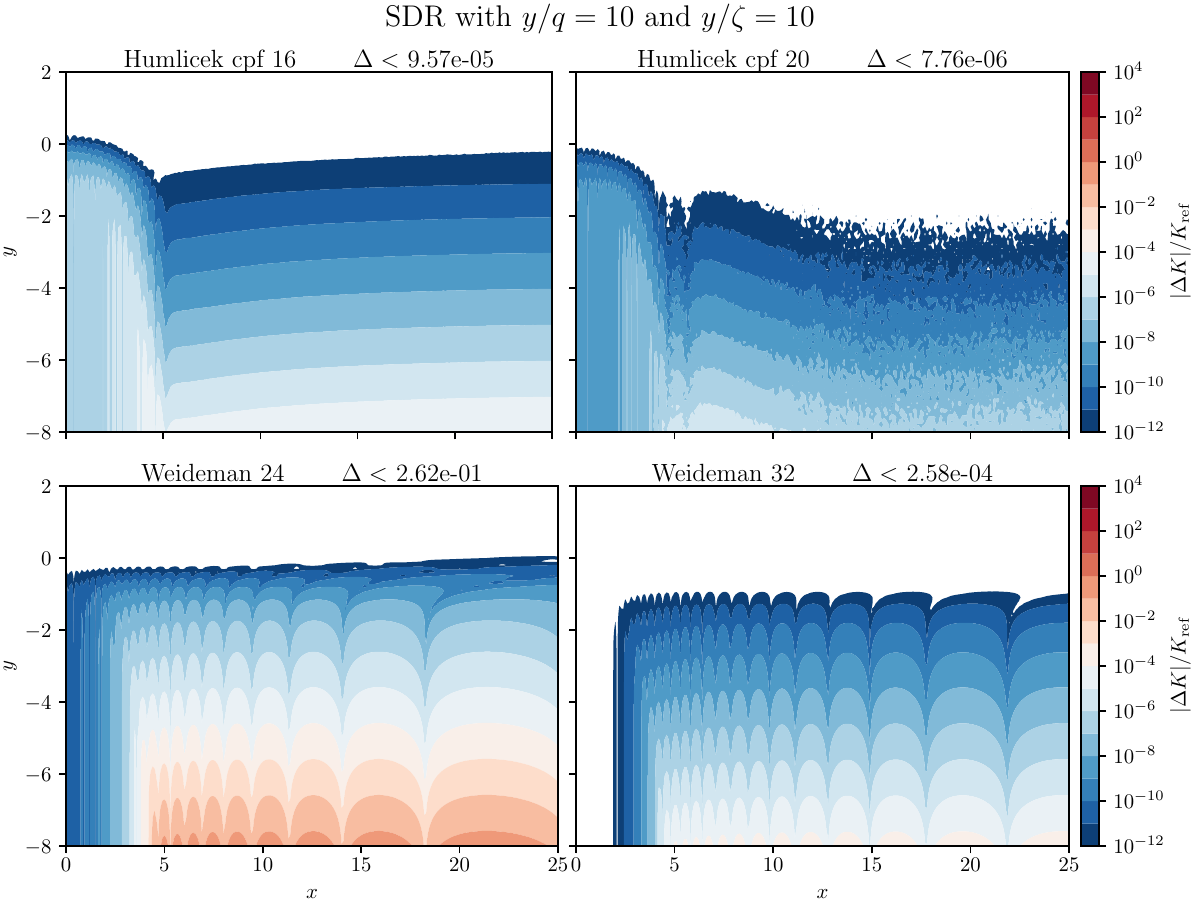}
 \caption{Contour plot of relative differences of the SDR function computed with the \protect\citeauthor{Humlicek79} (top) and \protect\citeauthor{Weideman94} complex error function codes compared to the \texttt{wofz} reference code.}
 \label{fig:sdr} 
\end{figure*}

The relative error of the SDR function computed with the \citet{Humlicek79} or \citet{Weideman94} code is shown in \qufig{fig:sdr}.
Interestingly the patterns are quite similar to those of \qufig{fig:cerf4sdv}.
In particular, for the Weideman $n=24$ rational approximation the errors can become significant already for moderately small $y$ (low pressure), whereas for 32 terms the approximation shows problems only for tiny $y$.
Both versions of the \citeauthor{Humlicek79} complex error function code give reliable results with four and five significant digits for the 16 and 20 term approximation, respectively.

\subsection{Performance}
\label{sec:speed}


For an assessment of the computational efficiency of various implementations we have used IPython's builtin ``magic function'' \texttt{\%timeit} similar to the tests described in \citet{Schreier18h}.
First we evaluate the SDV for the ``matrix'' of 251 $x$ grid points and 101 $y$ grid points used for all contour plots shown so far
(see \ref{app:timing}).
\qutab{tab:timeit_roots} shows that the times required for the three different versions of the complex square root \eqref{zpm} and the ``cancellation--safe'' alternative \eqref{zMinus} are approximately identical.
Both division and square root are classified as ``special function'' by \citet{Goedecker01} from a computational point of view, hence the costs of the \eqref{zPMbwb} and \eqref{zPMnew} versions should be roughly identical.

\begin{table}
 \caption{Execution time for various SDV implementations.
(All tests have been performed on a single Intel Core i5-9600 with 3.1\,GHz.)}
 \label{tab:timeit_roots}
 \begin{tabular}{ll}
  \hline
  function                 & time     \\
  \hline
  \texttt{cpf16}                           & 1.46\,ms \\
  \texttt{sdv\_real\_bwb} \eqref{zPMbwb}   & 4.37\,ms \\
  \texttt{sdv\_real\_casdv} \eqref{zPMnew} & 4.47\,ms \\
  \texttt{sdv\_naive} \eqref{zPMtnh}        & 4.91\,ms \\
  \texttt{sdv\_safe}  \eqref{zMinus}       & 4.94\,ms \\
  \hline
 \end{tabular}
\end{table}

Obviously these advanced line shapes are computationally more expensive.
For the following counts of floating-point operations only those required for every $x$ value (corresponding to a wavenumber grid point $\nu$) are considered, i.e.\ operations executed only once per line (e.g.\ division by $\gamma_\text{G}$) are ignored.
For both SDV and SDR two complex error functions have to be evaluated and in addition a complex square root ($z_+$ according to \eqref{sdrZ}) and a division ($z_-$ according to \eqref{zMinus}) is required,
a second division for the SDR \eqref{sdrFunction}, whereas for the Rautian function \eqref{rautian} there is a single extra division.
\citet{Tran13} have discussed the speed of the HT profile and concluded that
``the maximum ratio of the computer time needed for this calculation to that for the corresponding Voigt profile can thus reach $2 \times 2.5 = 5$.''
Note that for the HT function no further ``special function'' evaluations are required compared to the SDR.

For a first idea of the computational speed of the various line shapes we count the number of divisions and square roots, i.e.\ ``special function'' evaluations in the sense of \citet{Goedecker01}.
Assuming a single rational approximation for the complex error function (\citeauthor{Humlicek79}'s \texttt{cpf16} in the following) valid on the whole complex plane one division is required for the Voigt function.
Table \ref{tab:timeit_lineshapes} (second column) indicates that for the SDV and SDR functions a factor of four to five penalty compared to the Voigt function can be expected.

For the timings reported in \qutab{tab:timeit_lineshapes} (last column) a somewhat more realistic scenario is considered:
For line-by-line (lbl) molecular cross sections the contribution of a single spectral transition is usually computed in a finite interval with a fixed wavenumber cutoff (for example up to distances of $25 \rm\, cm^{-1}$ from the line center) or a cut at a multiple of the line half width.
Here a uniform $x$-grid from $0$ to $100 x_{1/2}$ and a grid point spacing defined by a fifth of the half width is used.
The results demonstrate that the Rautian function is only slightly slower than the Voigt function, and the SDV and SDR are roughly a factor two slower, in contrast to the estimate based on ``special function'' counts.

These results suggest that the additions and multiplications should also be considered.
In fact, with the Horner scheme 23 add-multiply operations are required to compute a Voigt function value with the \citet{Humlicek79} rational approximation generalized to 16 terms \citep{Schreier18h};
for the SDV and SDR the number of extra multiplications (in addition to the 46 multiplications for the two \texttt{cpf16} calls) is small (\qutab{tab:timeit_lineshapes}, third column).

\begin{table}
 \caption{Execution time for Voigt, Rautian, SDV, and SDR functions. The second and third columns give the number of ``special functions'' and multiplications required.
          The \texttt{cpf16} rational approximation of the complex error function is used for all functions.}
 \label{tab:timeit_lineshapes}
 \begin{tabular}{llll}
  \hline
  function     & sp.\ fct.   & mult. & time     \\   
  \hline
  Voigt        &  1  & 23    &  9.04\,ms \\  
  Rautian      &  2  & 24    &  9.49\,ms \\  
  SDV          &  4  & 49    &  19.4\,ms \\  
  SDR          &  5  & 50    &  19.8\,ms \\  
  \hline
 \end{tabular}
\end{table}

Finally we compute cross sections
\begin{equation}  \label{absXS}
 k(\nu,p,T) ~=~ \sum\limits_l S_l(T) \: g\Bigl(\nu-\hat\nu_l, \gamma_l^\text{(L)}(p,T), \gamma_l^\text{(G)}(T), \dots \Bigr)
\end{equation}
in the SWIR with line data (position $\hat\nu_l$, strength $S_l$ and broadening parameters) taken from the SEOM-IAS database.
The SWIR channels of TROPOMI \citep{Veefkind12} are exploited for remote sensing of atmospheric carbon monoxide (\chem{CO}) and methane (\chem{CH_4}).
For the retrieval of \chem{CO} vertical column densities with the Beer InfraRed Retrieval Algorithm (BIRRA) a spectral interval $4277 \, \text{--}\, 4303 \rm\,cm^{-1}$ is used \citep{GimenoGarcia11,Hochstaffl20s}, and in the following we report execution times for cross sections evaluated for a 20 level mid latitude summer atmosphere
(with top of atmosphere at 100\,km, \citep{Anderson86}).
For consistency with the continuum corrections and to allow a convolution of the monochromatic spectrum with the instrumental line shape function an extended spectral interval is considered.
The cross sections are computed using the \texttt{lbl2xs} function of the Py4CAtS package \citep[PYthon for Computational ATmospheric Spectroscopy,][]{Schreier19p}.

The results listed in \qutab{tab:time_lbl2xs} confirm that the Rautian line shape is slightly more expensive w.r.t.\ computational speed.
SDV and SDR times are approximately identical and larger than the Voigt times.
The modest time lag incurred by the speed-dependent profiles appears to be reasonable because only a small fraction of the lines has $\gamma_2$ and/or $\nu_\text{vc}$ defined,
i.e.\ the standard Voigt profile can be used for the majority of weak lines.

The high quality of the \citet{Humlicek79} 16-term rational approximation is not required in the line wings,
and further tests have therefore been conducted with a combination of this \texttt{cpf16} code and the \citet{Humlicek82} asymptotic rational approximation for large complex error function arguments $|z|>15$ 
(termed \texttt{hum1zpf16} in \citep{Schreier18h}).
This combined code provides a significant speed-up for all line shapes, with the relative performance (Voigt vs.\ speed-dependent profiles) being similar to the previous tests (\qutab{tab:time_lbl2xs} third block).

\begin{table}
 \caption{Execution time (seconds per line) for lbl molecular cross sections in the SWIR.
  In the first block the total number of lines in $4250 \, \text{--}\, 4330 \rm\,cm^{-1}$ and the number of lines with ``beyond Voigt'' parameters is given.
  For methane 62 lines have both non-zero speed-dependence and Dicke narrowing parameter.
  In the second block times for the \texttt{cpf16} complex error function code are listed, in the third block times for the \texttt{hum1cpf16} code.}
 \label{tab:time_lbl2xs}
 \begin{tabular}{lllll}
  \hline
  function             & \chem{CO} & \chem{CH_4} & \chem{H_2O} & total   \\
  \hline
  lines                & 117       & 6516        & 1262        & 7795    \\
  $\gamma_2>0$         & 10        & 637         & 39          &         \\
  $\nu_\text{vc}>0$    & 10        & 66          & 113         &         \\
  \hline
  Voigt                & 0.0521    & 0.0411      & 0.0417      & 0.0413 \\
  Rautian              & 0.0545    & 0.0436      & 0.0442      & 0.0438 \\
  SDV                  & 0.0580    & 0.0466      & 0.0437      & 0.0463 \\
  SDR                  & 0.0599    & 0.0462      & 0.0457      & 0.0463 \\
  \hline
  Voigt                & 0.0092    & 0.0068      & 0.0069      & 0.0069 \\
  Rautian              & 0.0121    & 0.0090      & 0.0090      & 0.0090 \\
  SDV                  & 0.0126    & 0.0097      & 0.0076      & 0.0094 \\
  SDR                  & 0.0157    & 0.0122      & 0.0100      & 0.0119 \\
  \hline
 \end{tabular}
\end{table}



\section{Summary}
\label{sec:conclusions}

Several line shapes have been recommended as a generalization of the standard Voigt function to model subtle broadening effects.
Various computational aspects of the SDV considering the speed-dependence of the line broadening had been discussed in a recent paper \citep{Schreier17}
and here we continue this study including the more advanced SDR that also models collisional narrowing effects.
First we present a brief survey of data available in the widely used HITRAN database \citep{Gordon17etal} and in the new SEOM-IAS database \citep{Birk17} including a discussion of the parameter range to be expected.
Next we demonstrate that for argument values possible in the upper atmosphere the SDV and SDR function values might suffer from cancellation errors;
these numerical problems can be avoided by a simple reformulation, thus eliminating the need for case selections and Taylor expansions etc.
Furthermore, the impact of various complex error function algorithms is studied:
the \citeauthor{Humlicek79} rational approximation generalized to 16 terms or the \citeauthor{Weideman94} 32-term approximation deliver SDV and SDR values with four or more significant digits.
(Note that in databases such as HITRAN, GEISA, or SEOM-IAS the line parameters are listed with a few digits accuracy only.)
Finally an assessment of the execution speed is provided.
First we discuss the number of floating point operations required for the various line shape functions: because of the extra divisions and square roots both SDV and SDR are computationally more expensive compared to the Voigt function.
However, for ``real world'' applications (molecular cross sections) the impact on the total code execution time is expected to be small.


\smallskip 
\section*{Acknowledgments}
This study has been motivated by a remark made by Frank Hase during a ``Workshop on SWIR Spectroscopy for S5p'' at the Karlsruhe Institute of Technology (KIT) end of January 2020 (Thanks to Claus Zehner for the invitation). 
FS is supported by the DFG project SCHR 1125/3-1;
PH is supported by the DAAD (German Academic Exchange Service).
Thanks to Thomas Trautmann for critical reading of the manuscript.
Thanks also to Alexandre Guillaume (JPL) for an interesting email discussion.

\begin{table*}[h]
 \caption{\label{tabNumbers}
 Numerical values for $y=10^{-8}$ and $q=10^{-9}$, hence $\alpha=8.5$ and $\delta\equiv Y = 2.5\cdot 10^{17}$.
 The first and fourth row correspond to the ``naive'' version, the second and fifth row to the ``safe'' version'' and the reference values in the third and sixth row are obtained by mpmath.}
 \tiny 
 \begin{tabular}{llllll}
  \hline
  $x$ & $\beta=\mathrm{Im}(X)$ & $\mathrm{Re}({\sqrt{X+Y}})$ & $\mathrm{Re}({\I z_-})$ & $\mathrm{Re}(w(iz_-))$ & $K_\text{sdv}$ \\
  \hline
  10 & 5000000000.0 & 500000000.0000001 & 1.1920928955078125e-07 & 6.829164078387363e-10 & 1.187268242909783e-10 \\
     & 5000000000.0 & 500000000.0000001 & 1.0849999999999996e-07 & 6.215659511546557e-10 & 5.737636760689775e-11 \\
     & 5000000000.0 & 500000000.0000001085  & 0.00000010849999999999994482775  & 6.2156585550296477893431e-10    & 5.73762544921658255221444e-11\\[1ex]   

  12 & 6000000000.0 & 500000000.0000001 & 1.1920928955078125e-07 & 4.720118241361795e-10 & -9.217775941157818e-11 \\
     & 6000000000.0 & 500000000.0000001 & 1.5249999999999992e-07 & 6.038271231846435e-10 & 3.9637539636885786e-11 \\
     & 6000000000.0 & 500000000.0000001525 & 0.00000015249999999999988890375 & 6.03827085623295707495819e-10  & 3.96375257362927384521548e-11  \\   
  \hline
 \end{tabular}
\end{table*}

\begin{footnotesize}
\bibliography{JOURNALS,atmos,comp,math,molec,planets,radiation,remote_sensing,voigt}

\begin{thebibliography}{45}
\providecommand{\natexlab}[1]{#1}
\providecommand{\url}[1]{\texttt{#1}}
\expandafter\ifx\csname urlstyle\endcsname\relax
  \providecommand{\doi}[1]{doi: #1}\else
  \providecommand{\doi}{doi: \begingroup \urlstyle{rm}\Url}\fi

\bibitem[Armstrong(1967)]{Armstrong67}
B.H. Armstrong.
\newblock Spectrum line profiles: The {Voigt} function.
\newblock \emph{J.\ Quant.\ Spectrosc.\ \& Radiat.\ Transfer}, 7:\penalty0
  61--88, 1967.
\newblock \doi{10.1016/0022-4073(67)90057-X}.

\bibitem[Dicke(1953)]{Dicke53}
R.~H. Dicke.
\newblock The effect of collisions upon the {Doppler} width of spectral lines.
\newblock \emph{Phys.\ Rev.}, 89:\penalty0 472--473, 1953.
\newblock \doi{10.1103/PhysRev.89.472}.

\bibitem[Varghese and Hanson(1984)]{Varghese84}
P.L. Varghese and R.K. Hanson.
\newblock Collisional narrowing effects on spectral line shapes measured at
  high resolution.
\newblock \emph{Appl.\ Opt.}, 23\penalty0 (14):\penalty0 2376--2385, 1984.
\newblock \doi{10.1364/AO.23.002376}.

\bibitem[Hartmann et~al.(2008)Hartmann, Boulet, and Robert]{Hartmann08}
J.M. Hartmann, C.~Boulet, and D.~Robert.
\newblock \emph{Collisional Effects on Molecular Spectra}.
\newblock Elsevier, 2008.

\bibitem[Tennyson et~al.(2014)Tennyson, Bernath, Campargue, Cs\'{a}sz\'{a}r,
  Daumont, Gamache, Hodges, Lisak, Naumenko, Rothman, Tran, Zobov, Buldyreva,
  Boone, Vizia, Gianfrani, Hartmann, McPheat, Weidmann, Murray, Ngo, and
  Polyansky]{Tennyson14}
J.~Tennyson, P.F. Bernath, A.~Campargue, A.G. Cs\'{a}sz\'{a}r, L.~Daumont, R.R.
  Gamache, J.T. Hodges, D.~Lisak, O.V. Naumenko, L.S. Rothman, H.~Tran, N.F.
  Zobov, J.~Buldyreva, C.D. Boone, M.D.~De Vizia, L.~Gianfrani, J.-M. Hartmann,
  R.~McPheat, D.~Weidmann, J.~Murray, N.H. Ngo, and O.L. Polyansky.
\newblock Recommended isolated-line profile for representing high-resolution
  spectroscopic transitions ({IUPAC} technical report).
\newblock \emph{Pure Appl.\ Chem.}, 86\penalty0 (12):\penalty0 1931--1943,
  2014.
\newblock \doi{10.1515/pac-2014-0208}.

\bibitem[Tran et~al.(2013)Tran, Ngo, and Hartmann]{Tran13}
H.~Tran, N.H. Ngo, and J.-M. Hartmann.
\newblock Efficient computation of some speed-dependent isolated line profiles.
\newblock \emph{J.\ Quant.\ Spectrosc.\ \& Radiat.\ Transfer}, 129:\penalty0
  199 -- 203, 2013.
\newblock \doi{10.1016/j.jqsrt.2013.06.015}.
\newblock Erratum: JQSRT 134, 104 (2014).

\bibitem[Schneider and Hase(2009)]{Schneider09b}
M.~Schneider and F.~Hase.
\newblock Improving spectroscopic line parameters by means of atmospheric
  spectra: Theory and example for water vapor and solar absorption spectra.
\newblock \emph{J.\ Quant.\ Spectrosc.\ \& Radiat.\ Transfer}, 110\penalty0
  (17):\penalty0 1825 -- 1839, 2009.
\newblock \doi{10.1016/j.jqsrt.2009.04.011}.

\bibitem[Schneider et~al.(2011)Schneider, Hase, Blavier, Toon, and
  Leblanc]{Schneider11}
M.~Schneider, F.~Hase, J.-F. Blavier, G.C. Toon, and T.~Leblanc.
\newblock An empirical study on the importance of a speed-dependent {Voigt}
  line shape model for tropospheric water vapor profile remote sensing.
\newblock \emph{J.\ Quant.\ Spectrosc.\ \& Radiat.\ Transfer}, 112\penalty0
  (3):\penalty0 465--474, 2011.
\newblock \doi{10.1016/j.jqsrt.2010.09.008}.

\bibitem[Boone et~al.(2007)Boone, Walker, and Bernath]{Boone07}
C.D. Boone, K.A. Walker, and P.F. Bernath.
\newblock Speed--dependent {Voigt} profile for water vapor in infrared remote
  sensing applications.
\newblock \emph{J.\ Quant.\ Spectrosc.\ \& Radiat.\ Transfer}, 105:\penalty0
  525--532, 2007.
\newblock \doi{10.1016/j.jqsrt.2006.11.015}.

\bibitem[Kuze et~al.(2009)Kuze, Suto, Nakajima, and Hamazaki]{Kuze09}
A.~Kuze, H.~Suto, M.~Nakajima, and T.~Hamazaki.
\newblock Thermal and near infrared sensor for carbon observation
  {Fourier}-transform spectrometer on the {Greenhouse Gases Observing
  Satellite} for greenhouse gases monitoring.
\newblock \emph{Appl.\ Opt.}, 48\penalty0 (35):\penalty0 6716--6733, 2009.
\newblock \doi{10.1364/AO.48.006716}.

\bibitem[Crisp et~al.(2004)Crisp, Atlas, Breon, Brown, Burrows, Ciais, Connor,
  Doney, Fung, Jacob, Miller, O'Brien, Pawson, Randerson, Rayner, Salawitch,
  Sander, Sen, Stephens, Tans, Toon, Wennberg, Wofsy, Yung, Kuang, Chudasama,
  Sprague, Weiss, Pollock, Kenyon, and Schroll]{Crisp04}
D.~Crisp, R.M. Atlas, F.-M. Breon, L.R. Brown, J.P. Burrows, P.~Ciais, B.J.
  Connor, S.C. Doney, I.Y. Fung, D.J. Jacob, C.E. Miller, D.~O'Brien,
  S.~Pawson, J.T. Randerson, P.~Rayner, R.J. Salawitch, S.P. Sander, B.~Sen,
  G.L. Stephens, P.P. Tans, G.C. Toon, P.O. Wennberg, S.C. Wofsy, Y.L. Yung,
  Z.~Kuang, B.~Chudasama, G.~Sprague, B.~Weiss, R.~Pollock, D.~Kenyon, and
  S.~Schroll.
\newblock The orbiting carbon observatory ({OCO}) mission.
\newblock \emph{Adv.\ Space Res.}, 34\penalty0 (4):\penalty0 700 -- 709, 2004.
\newblock \doi{10.1016/j.asr.2003.08.062}.

\bibitem[Veefkind et~al.(2012)Veefkind, Aben, McMullan, F{\"o}rster, de~Vries,
  Otter, Claas, Eskes, de~Haan, Kleipool, van Weele, Hasekamp, Hoogeveen,
  Landgraf, Snel, Tol, Ingmann, Voors, Kruizinga, Vink, Visser, and
  Levelt]{Veefkind12}
J.P. Veefkind, I.~Aben, K.~McMullan, H.~F{\"o}rster, J.~de~Vries, G.~Otter,
  J.~Claas, H.J. Eskes, J.F. de~Haan, Q.~Kleipool, M.~van Weele, O.~Hasekamp,
  R.~Hoogeveen, J.~Landgraf, R.~Snel, P.~Tol, P.~Ingmann, R.~Voors,
  B.~Kruizinga, R.~Vink, H.~Visser, and P.F. Levelt.
\newblock {TROPOMI} on the {ESA} {Sentinel-5 Precursor}: A {GMES} mission for
  global observations of the atmospheric composition for climate, air quality
  and ozone layer applications.
\newblock \emph{Remote Sensing of Environment}, 120:\penalty0 70 -- 83, 2012.
\newblock \doi{10.1016/j.rse.2011.09.027}.
\newblock The Sentinel Missions - New Opportunities for Science.

\bibitem[Nikitin et~al.(2010)Nikitin, Lyulin, Mikhailenko, Perevalov, Filippov,
  Grigoriev, Morino, Yokota, Kumazawa, and Watanabe]{Nikitin10}
A.V. Nikitin, O.M. Lyulin, S.N. Mikhailenko, V.I. Perevalov, N.N. Filippov,
  I.M. Grigoriev, I.~Morino, T.~Yokota, R.~Kumazawa, and T.~Watanabe.
\newblock {GOSAT}-2009 methane spectral line list in the 5550 --
  $6236\rm\,cm^{-1}$ range.
\newblock \emph{J.\ Quant.\ Spectrosc.\ \& Radiat.\ Transfer}, 111\penalty0
  (12-13):\penalty0 2211--2224, 2010.
\newblock \doi{10.1016/j.jqsrt.2010.05.010}.

\bibitem[Nikitin et~al.(2015)Nikitin, Lyulin, Mikhailenko, Perevalov, Filippov,
  Grigoriev, Morino, Yoshida, and Matsunaga]{Nikitin15}
A.V. Nikitin, O.M. Lyulin, S.N. Mikhailenko, V.I. Perevalov, N.N. Filippov,
  I.M. Grigoriev, I.~Morino, Y.~Yoshida, and T.~Matsunaga.
\newblock {GOSAT}-2014 methane spectral line list.
\newblock \emph{J.\ Quant.\ Spectrosc.\ \& Radiat.\ Transfer}, 154:\penalty0 63
  -- 71, 2015.
\newblock \doi{10.1016/j.jqsrt.2014.12.003}.

\bibitem[Oyafuso et~al.(2017)Oyafuso, Payne, Drouin, Devi, Benner, Sung, Yu,
  Gordon, Kochanov, Tan, Crisp, Mlawer, and Guillaume]{Oyafuso17}
F.~Oyafuso, V.H. Payne, B.J. Drouin, V.M. Devi, D.C. Benner, K.~Sung, S.~Yu,
  I.E. Gordon, R.~Kochanov, Y.~Tan, D.~Crisp, E.J. Mlawer, and A.~Guillaume.
\newblock High accuracy absorption coefficients for the {Orbiting Carbon
  Observatory-2 (OCO-2)} mission: Validation of updated carbon dioxide
  cross-sections using atmospheric spectra.
\newblock \emph{J.\ Quant.\ Spectrosc.\ \& Radiat.\ Transfer}, 203:\penalty0
  213 -- 223, 2017.
\newblock \doi{10.1016/j.jqsrt.2017.06.012}.
\newblock HITRAN2016 Special Issue.

\bibitem[Galli et~al.(2014)Galli, Guerlet, Butz, Aben, Suto, Kuze, Deutscher,
  Notholt, Wunch, Wennberg, Griffith, Hasekamp, and Landgraf]{Galli14}
A.~Galli, S.~Guerlet, A.~Butz, I.~Aben, H.~Suto, A.~Kuze, N.~M. Deutscher,
  J.~Notholt, D.~Wunch, P.~O. Wennberg, D.~W.~T. Griffith, O.~Hasekamp, and
  J.~Landgraf.
\newblock The impact of spectral resolution on satellite retrieval accuracy of
  \chem{CO_2} and \chem{CH_4}.
\newblock \emph{Atmos.\ Meas.\ Tech.}, 7\penalty0 (4):\penalty0 1105--1119,
  2014.
\newblock \doi{10.5194/amt-7-1105-2014}.

\bibitem[Checa-Garcia et~al.(2015)Checa-Garcia, Landgraf, Galli, Hase, Velazco,
  Tran, Boudon, Alkemade, and Butz]{ChecaGarcia15}
R.~Checa-Garcia, J.~Landgraf, A.~Galli, F.~Hase, V.A. Velazco, H.~Tran,
  V.~Boudon, F.~Alkemade, and A.~Butz.
\newblock Mapping spectroscopic uncertainties into prospective methane
  retrieval errors from {Sentinel-5} and its precursor.
\newblock \emph{Atmos.\ Meas.\ Tech.}, 8\penalty0 (9):\penalty0 3617--3629,
  2015.
\newblock \doi{10.5194/amt-8-3617-2015}.

\bibitem[Hochstaffl and Schreier(2020{\natexlab{a}})]{Hochstaffl20s}
P.~Hochstaffl and F.~Schreier.
\newblock Impact of molecular spectroscopy on carbon monoxide abundances from
  {SCIAMACHY}.
\newblock \emph{Remote Sensing}, 12\penalty0 (7):\penalty0 1084,
  2020{\natexlab{a}}.
\newblock \doi{10.3390/rs12071084}.

\bibitem[Hochstaffl and Schreier(2020{\natexlab{b}})]{Hochstaffl20t}
P.~Hochstaffl and F.~Schreier.
\newblock Impact of molecular spectroscopy on carbon monoxide abundances from
  {TROPOMI}.
\newblock \emph{Remote Sensing}, October 2020{\natexlab{b}}.
\newblock Manuscript accepted subject to minor revisions.

\bibitem[Birk et~al.(2017)Birk, Wagner, Loos, Mondelain, and Campargue]{Birk17}
M.~Birk, G.~Wagner, J.~Loos, D.~Mondelain, and A.~Campargue.
\newblock {ESA SEOM--IAS} - spectroscopic parameters database $2.3\rm\,\mu m$
  region [data set].
\newblock Zenodo, 2017.
\newblock URL \url{http://doi.org/10.5281/zenodo.1009126}.

\bibitem[Gordon et~al.(2017)Gordon, Rothman, et~al.]{Gordon17etal}
I.E. Gordon, L.S. Rothman, et~al.
\newblock The {HITRAN2016} molecular spectroscopic database.
\newblock \emph{J.\ Quant.\ Spectrosc.\ \& Radiat.\ Transfer}, 203:\penalty0 3
  -- 69, 2017.
\newblock \doi{10.1016/j.jqsrt.2017.06.038}.

\bibitem[Schreier(2017)]{Schreier17}
F.~Schreier.
\newblock Computational aspects of speed-dependent {Voigt} profiles.
\newblock \emph{J.\ Quant.\ Spectrosc.\ \& Radiat.\ Transfer}, 187:\penalty0
  44--53, 2017.
\newblock \doi{10.1016/j.jqsrt.2016.08.009}.

\bibitem[Schreier(2011)]{Schreier11v}
F.~Schreier.
\newblock Optimized implementations of rational approximations for the {Voigt}
  and complex error function.
\newblock \emph{J.\ Quant.\ Spectrosc.\ \& Radiat.\ Transfer}, 112\penalty0
  (6):\penalty0 1010--1025, 2011.
\newblock \doi{10.1016/j.jqsrt.2010.12.010}.

\bibitem[Schreier(2018)]{Schreier18h}
F.~Schreier.
\newblock The {Voigt} and complex error function: {Huml\'\i\v{c}ek}'s rational
  approximation generalized.
\newblock \emph{Mon.\ Not.\ Roy.\ Astron.\ Soc.}, 479\penalty0 (3):\penalty0
  3068--3075, 2018.
\newblock \doi{10.1093/mnras/sty1680}.

\bibitem[Loos et~al.(2017{\natexlab{a}})Loos, Birk, and Wagner]{Loos17a}
J.~Loos, M.~Birk, and G.~Wagner.
\newblock Measurement of air-broadening line shape parameters and temperature
  dependence parameters of \chem{H_2O} lines in the spectral ranges $1850
  \mbox{--} 2280 \rm\, cm^{-1}$ and $2390 \mbox{--} 4000\rm\, cm^{-1}$.
\newblock \emph{J.\ Quant.\ Spectrosc.\ \& Radiat.\ Transfer}, 203:\penalty0
  103 -- 118, 2017{\natexlab{a}}.
\newblock \doi{10.1016/j.jqsrt.2017.03.033}.

\bibitem[Loos et~al.(2017{\natexlab{b}})Loos, Birk, and Wagner]{Loos17p}
J.~Loos, M.~Birk, and G.~Wagner.
\newblock Measurement of positions, intensities and self-broadening line shape
  parameters of \chem{H_2O} lines in the spectral ranges $1850 \mbox{--} 2280
  \rm\, cm^{-1}$ and $2390 \mbox{--} 4000\rm\, cm^{-1}$.
\newblock \emph{J.\ Quant.\ Spectrosc.\ \& Radiat.\ Transfer}, 203:\penalty0
  119 -- 132, 2017{\natexlab{b}}.
\newblock \doi{10.1016/j.jqsrt.2017.02.013}.

\bibitem[Wcislo et~al.(2016)Wcislo, Gordon, Tran, Tan, Hu, Campargue, Kassi,
  Romanini, Hill, Kochanov, and Rothman]{Wcislo16}
P.~Wcislo, I.E. Gordon, H.~Tran, Y.~Tan, S.-M. Hu, A.~Campargue, S.~Kassi,
  D.~Romanini, C.~Hill, R.V. Kochanov, and L.S. Rothman.
\newblock The implementation of non-{Voigt} line profiles in the {HITRAN}
  database: \chem{H_2} case study.
\newblock \emph{J.\ Quant.\ Spectrosc.\ \& Radiat.\ Transfer}, 177:\penalty0
  175--191, 2016.
\newblock \doi{10.1016/j.jqsrt.2016.01.024}.

\bibitem[Gamache and Vispoel(2018)]{Gamache20}
R.R. Gamache and B.~Vispoel.
\newblock On the temperature dependence of half-widths and line shifts for
  molecular transitions in the microwave and infrared regions.
\newblock \emph{J.\ Quant.\ Spectrosc.\ \& Radiat.\ Transfer}, 217:\penalty0
  440 -- 452, 2018.
\newblock \doi{10.1016/j.jqsrt.2018.05.019}.

\bibitem[Stolarczyk et~al.(2020)Stolarczyk, Thibault, Cybulski, J\'o\'zwiak,
  Kowzan, Vispoel, Gordon, Rothman, Gamache, and Wcis{\l}o]{Stolarczyk20}
N.~Stolarczyk, F.~Thibault, H.~Cybulski, H.~J\'o\'zwiak, G.~Kowzan, B.~Vispoel,
  I.E. Gordon, L.S. Rothman, R.R. Gamache, and P.~Wcis{\l}o.
\newblock Evaluation of different parameterizations of temperature dependences
  of the line-shape parameters based on ab initio calculations: Case study for
  the {HITRAN} database.
\newblock \emph{J.\ Quant.\ Spectrosc.\ \& Radiat.\ Transfer}, 240:\penalty0
  106676, 2020.
\newblock \doi{10.1016/j.jqsrt.2019.106676}.

\bibitem[Abramowitz and Stegun(1964)]{AbSt64}
M.~Abramowitz and I.A. Stegun.
\newblock \emph{Handbook of Mathematical Functions}.
\newblock National Bureau of Standards, AMS55, New York, 1964.

\bibitem[Olver et~al.(2010)Olver, Lozier, Boisvert, and Clark]{NIST-handbook}
F.W.J. Olver, D.W. Lozier, R.F. Boisvert, and C.W. Clark, editors.
\newblock \emph{{NIST} Handbook of Mathematical Functions}.
\newblock Cambridge University Press, New York, NY, 2010.
\newblock Print companion to \cite{DLMF}.

\bibitem[DLMF()]{DLMF}
DLMF.
\newblock \emph{{NIST} Digital Library of Mathematical Functions}.
\newblock National Institute of Standards and Technology.
\newblock URL \url{http://dlmf.nist.gov/}.
\newblock Online companion to \cite{NIST-handbook}.

\bibitem[Boone et~al.(2011)Boone, Walker, and Bernath]{Boone11}
C.D. Boone, K.A. Walker, and P.F. Bernath.
\newblock An efficient analytical approach for calculating line mixing in
  atmospheric remote sensing applications.
\newblock \emph{J.\ Quant.\ Spectrosc.\ \& Radiat.\ Transfer}, 112\penalty0
  (6):\penalty0 980 -- 989, 2011.
\newblock \doi{10.1016/j.jqsrt.2010.11.013}.

\bibitem[Huml\'\i\v{c}ek(1979)]{Humlicek79}
J.~Huml\'\i\v{c}ek.
\newblock An efficient method for evaluation of the complex probability
  function: the {Voigt} function and its derivatives.
\newblock \emph{J.\ Quant.\ Spectrosc.\ \& Radiat.\ Transfer}, 21:\penalty0
  309--313, 1979.
\newblock \doi{10.1016/0022-4073(79)90062-1}.

\bibitem[Huml\'\i\v{c}ek(1982)]{Humlicek82}
J.~Huml\'\i\v{c}ek.
\newblock Optimized computation of the {Voigt} and complex probability
  function.
\newblock \emph{J.\ Quant.\ Spectrosc.\ \& Radiat.\ Transfer}, 27:\penalty0
  437--444, 1982.
\newblock \doi{10.1016/0022-4073(82)90078-4}.

\bibitem[Weideman(1994)]{Weideman94}
J.A.C. Weideman.
\newblock Computation of the complex error function.
\newblock \emph{SIAM J.\ Num.\ Anal.}, 31:\penalty0 1497--1518, 1994.
\newblock \doi{10.1137/0731077}.

\bibitem[Johansson et~al.(2013)]{mpmath}
Fredrik Johansson et~al.
\newblock \emph{mpmath: a {P}ython library for arbitrary-precision
  floating-point arithmetic (version 0.18)}, December 2013.
\newblock {\tt http://mpmath.org/}.

\bibitem[Panchekha et~al.(2015)Panchekha, Sanchez-Stern, Wilcox, and
  Tatlock]{Panchekha15}
P.~Panchekha, A.~Sanchez-Stern, J.R. Wilcox, and Z.~Tatlock.
\newblock Automatically improving accuracy for floating point expressions.
\newblock In \emph{PLDI '15: Proceedings of the 36th ACM SIGPLAN Conference on
  Programming Language Design and Implementation}. Association for Computing
  Machinery, 2015.
\newblock \doi{10.1145/10.1145/2737924.2737959}.
\newblock URL \url{https://herbie.uwplse.org/pldi15-paper.pdf}.

\bibitem[Johnson(Last access August 2020)]{faddeeva}
S.G. Johnson.
\newblock Faddeeva package, Last access August 2020.
\newblock URL \url{http://ab-initio.mit.edu/Faddeeva}.

\bibitem[Johnson and Wuttke(Last access August 2020)]{libcerf}
S.G. Johnson and J.~Wuttke.
\newblock libcerf, numeric library for complex error functions, Last access
  August 2020.
\newblock URL \url{https://jugit.fz-juelich.de/mlz/libcerf}.

\bibitem[Whiting(1968)]{Whiting68}
E.E. Whiting.
\newblock An empirical approximation to the {Voigt} profile.
\newblock \emph{J.\ Quant.\ Spectrosc.\ \& Radiat.\ Transfer}, 8:\penalty0
  1379--1384, 1968.
\newblock \doi{10.1016/0022-4073(68)90081-2}.

\bibitem[Goedecker and Hoisie(2001)]{Goedecker01}
Stefan Goedecker and Adolfy Hoisie.
\newblock \emph{Performance Optimization of Numerically Intensive Codes}.
\newblock SIAM, Philadelphia, PA, 2001.

\bibitem[{Gimeno Garc\'{\i}a} et~al.(2011){Gimeno Garc\'{\i}a}, Schreier,
  Lichtenberg, and Slijkhuis]{GimenoGarcia11}
S.~{Gimeno Garc\'{\i}a}, F.~Schreier, G.~Lichtenberg, and S.~Slijkhuis.
\newblock Near infrared nadir retrieval of vertical column densities:
  methodology and application to {SCIAMACHY}.
\newblock \emph{Atmos.\ Meas.\ Tech.}, 4\penalty0 (12):\penalty0 2633--2657,
  2011.
\newblock \doi{10.5194/amt-4-2633-2011}.

\bibitem[Anderson et~al.(1986)Anderson, Clough, Kneizys, Chetwynd, and
  Shettle]{Anderson86}
G.P. Anderson, S.A. Clough, F.X. Kneizys, J.H. Chetwynd, and E.P. Shettle.
\newblock {AFGL} atmospheric constituent profiles (0 -- $\rm 120\,km$).
\newblock Technical Report TR-86-0110, AFGL, 1986.

\bibitem[Schreier et~al.(2019)Schreier, {Gimeno Garc{\'\i}a}, Hochstaffl, and
  St\"adt]{Schreier19p}
F.~Schreier, S.~{Gimeno Garc{\'\i}a}, P.~Hochstaffl, and S.~St\"adt.
\newblock {Py4CAtS} --- {PYthon} for {Computational} {ATmospheric}
  {Spectroscopy}.
\newblock \emph{Atmosphere}, 10\penalty0 (5):\penalty0 262, 2019.
\newblock \doi{10.3390/atmos10050262}.

\end{thebibliography}
\end{footnotesize}

\appendix

\setcounter{figure}{0}

\section{Example of the SDV calculation for small values of \texorpdfstring{$y$ and $q$}{y and q}}

To further illustrate the discussion of subsection \ref{sec:yTiny} a comparsion for two values of $x$ and for $y=10^{-8}$ and $q=0.1 y$ is given in \qutab{tabNumbers}.
The numbers shown here are copied from the (I)Python shell output without any formatting of the \texttt{print} statement.
Note that for the mpmath calculations shown here a ``triple precision'' has been exploited (\texttt{mp.dps=24}), in contrast to the quadruple precision used for \qufig{fig:sdv_yTiny}.

\section{Code snippets used for time benchmarks}
\label{app:timing}

For the first test assessing the performance of the various ways to evaluate the complex square root and the difference thereof the SDV has been computed on the ``matrix'' of $x$ and $y$ values used for all contour plots (note that Python is case sensitive):
\begin{small}
\begin{verbatim}
In [1]: from sdv import *
In [2]: # standard Voigt variables
    ...: xx=np.linspace(0.,25.,251)  # linear grid
    ...: yy=np.logspace(-8,2,101)    # logarithmic grid
    ...: # kind of 'matrices' required for contour plots
    ...: XX, YY = np.meshgrid(xx, yy)
    ...: ZZ = XX +1j*YY
In [3]: %timeit sdv_safe(XX, YY, 0.1*YY, cpf16p)
\end{verbatim}
\end{small}

In the second test the SDR has been calculated for the 101 $y$ values as defined above with an equidistant $x$ grid in the interval defined by the Voigt function half width:
\begin{small}
\begin{verbatim}
xx=np.linspace(0.,25.,251)
yy=np.logspace(-8,2,101)
%%timeit
for y in yy:
    xHalf = whiting(y); xx=arange(0.0,100*xHalf,xHalf/5)
    sdr(xx, y, 0.1*y, 0.1*y, cpf16p)
\end{verbatim}
\end{small}


\end{document}